\documentclass{mem}
\usepackage{natbib}\usepackage{txfonts}\usepackage{balance}
\usepackage{graphicx}
\usepackage[a4paper]{hyperref}
\idline{00}{000}
\begin{document}
\def\teff{$T\rm_{eff }$}
\def\kms{$\mathrm {km s}^{-1}$}

\title{
X-ray activity cycles in stellar coronae
}

   \subtitle{}

\author{
J.  Robrade
\and J.H.M.M. Schmitt
\and A. Hempelmann
          }

%  \offprints{J. Robrade}

\institute{
Universit\"at Hamburg,
Hamburger Sternwarte, Gojenbergsweg 112
D-21029 Hamburg, Germany;
\email{jrobrade@hs.uni-hamburg.de}
}

\authorrunning{Robrade et al.}

\titlerunning{X-ray activity cycles in stellar coronae}

\abstract{
We present updated results from the ongoing  XMM-Newton monitoring program of moderately active, solar-like stars to investigate stellar X-ray activity cycles;
here we report on the binary systems $\alpha$\,Cen~A/B and 61\,Cyg~A/B. 
For 61\,Cyg~A we find a coronal X-ray cycle which clearly reflects the
chromospheric activity cycle and is in phase with a ROSAT campaign performed in the 1990s.
61\,Cyg~A is the first example of a persistent coronal cycle observed on a star other than the Sun.
The changes of its coronal properties during the cycle resemble the solar behaviour.
The coronal activity of 61\,Cyg~B is more irregular, but also follows the chromospheric activity.
Long-term variability is also present on $\alpha$\,Cen~A and B.
We find that $\alpha$\,Cen~A, a G2 star very similar to our Sun, fainted in X-rays by at least 
an order of magnitude during the observation program. This behaviour has never been observed before on
$\alpha$\,Cen~A, but is rather similar to the X-ray behaviour of the Sun.
The X-ray emission of the $\alpha$\,Cen system is dominated in our observations by $\alpha$\,Cen B, which might also have an activity cycle
indicated by a significant fainting since 2005.

}
\maketitle{}

\section{Introduction}

Chromospheric stellar activity cycles in analogy to the 11 year solar activity cycle are well known 
from the Mt. Wilson monitoring of CaII H+K emission (S-index) \citep{bal95}.
In contrast, coronal cycles are known only for very few examples, mainly because
of lacking X-ray data covering typical stellar cycles of several of years.
In order to study possible coronal activity cycles of solar-like stars we initiated a long-term X-ray monitoring
program with XMM-Newton.

$\alpha$\,Centauri is the nearest stellar system at a distance of 1.3\,pc, with $\alpha$~Cen A/B
consisting of a G2V~(A) and a K1V~(B) separated by roughly 25~AU and an orbital period of 80~years.
The age of the system is thought to be slightly larger than that of the Sun, correspondingly
both stars are slow rotators with a rather inactive corona. 
$\alpha$\,Cen\,A is a nearly perfect solar twin, which 
raises the question whether an activity cycle as observed on the Sun is also present.  
The system has already been
spatially resolved in X-rays with {\it Einstein} \citep{gol82}, ROSAT and recently {\it Chandra}.
The X-ray luminosity of the system is dominated in all observations by $\alpha$\,Cen\,B, 
which was a factor 2--3 brighter than $\alpha$\,Cen~A at energies above 0.2\,keV.
High resolution X-ray spectra \citep{ras03}
revealed similar, solar-like coronal properties for both components with
$\alpha$\,Cen~B being slightly hotter and dominant above 1.5\,MK.

61~Cygni, a K5V (A) and a K7V (B) star at a distance of 3.5\,pc was part of the Mt. Wilson program.
61\,Cyg~A exhibits a very regular chromospheric activity cycle with a period of 7.3 years and a 
nearly symmetric rise and decay phase; in contrast the 11.7 year cycle of 61\,Cyg~B is more irregular, asymmetric 
and its mean activity index is higher. \cite{hem03} analysed 4.5 years of ROSAT HRI data taken in the 1990s
and find that long-term X-ray variability correlates with chromospheric activity for both stars.

First XMM-Newton results were presented  for $\alpha$\,Cen by \cite{rob05} and for 61\,Cyg by \cite{hem06}.
Similarly, \cite{fav04} presented the data of the more active G2~star HD\,81809 (P=8.2\,yr) and find clear evidence for large amplitude 
X-ray variability in phase with the known chromospheric activity cycle.

\section{Observations and data analysis}

61\,Cyg  and $\alpha$\,Cen were each observed twice a year in the course of the long-term monitoring program 
of moderately active stars with XMM-Newton.
Over the last years we obtained snapshot like exposures of 5\,--\,15\,ks each and
present data from 10~(61\,Cyg) and 8~($\alpha$\,Cen) observations,
which allow us to study variability on timescales of several years and investigate
possible cyclic behaviour. For this analysis we considered only data from 
their quasi-quiescent states by excluding time periods of enhanced activity or flares 
on the basis of derived light curves. 
We use for $\alpha$\,Cen (sep. 10-12\,\arcsec) a PSF (Point Spread Function) fitting algorithm which is applied to the event distribution 
in the sky-plane, for 61\,Cyg (sep. $>$30\,\arcsec) we use individual extraction regions.
X-ray luminosities, temperatures and emission measures (EM) were derived for the individual targets and exposures from EPIC, i.e. MOS and PN, data.
Spectral analysis uses multi-temperature VAPEC models in the energy range 0.2\,--\,5.0\,keV,
however sufficient signal is mostly present only up to 2.0\,keV. 

\section{61 Cygni} 

\begin{figure}[]
\includegraphics[width=32mm]{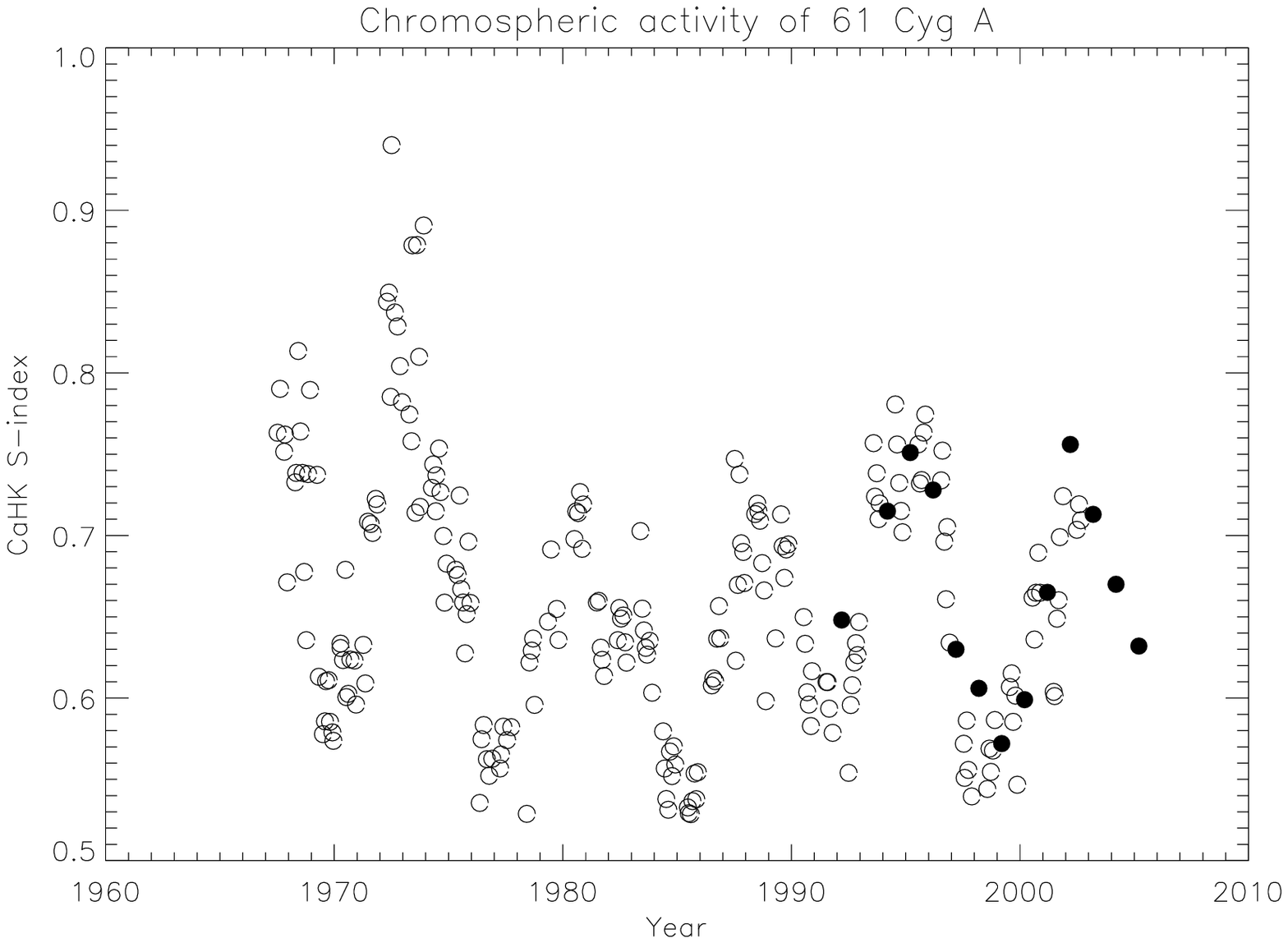}
\includegraphics[width=32mm]{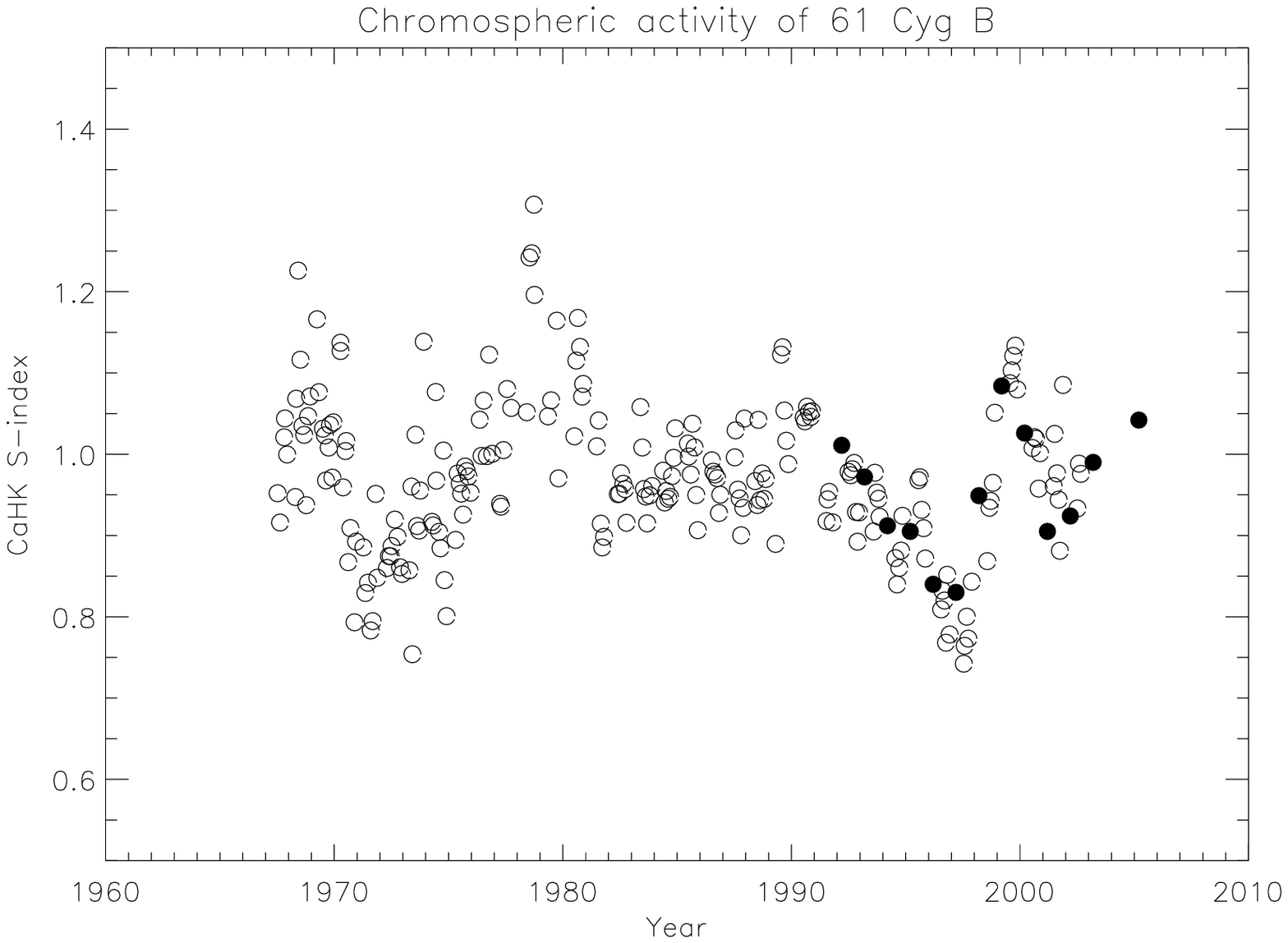}

\includegraphics[width=32mm]{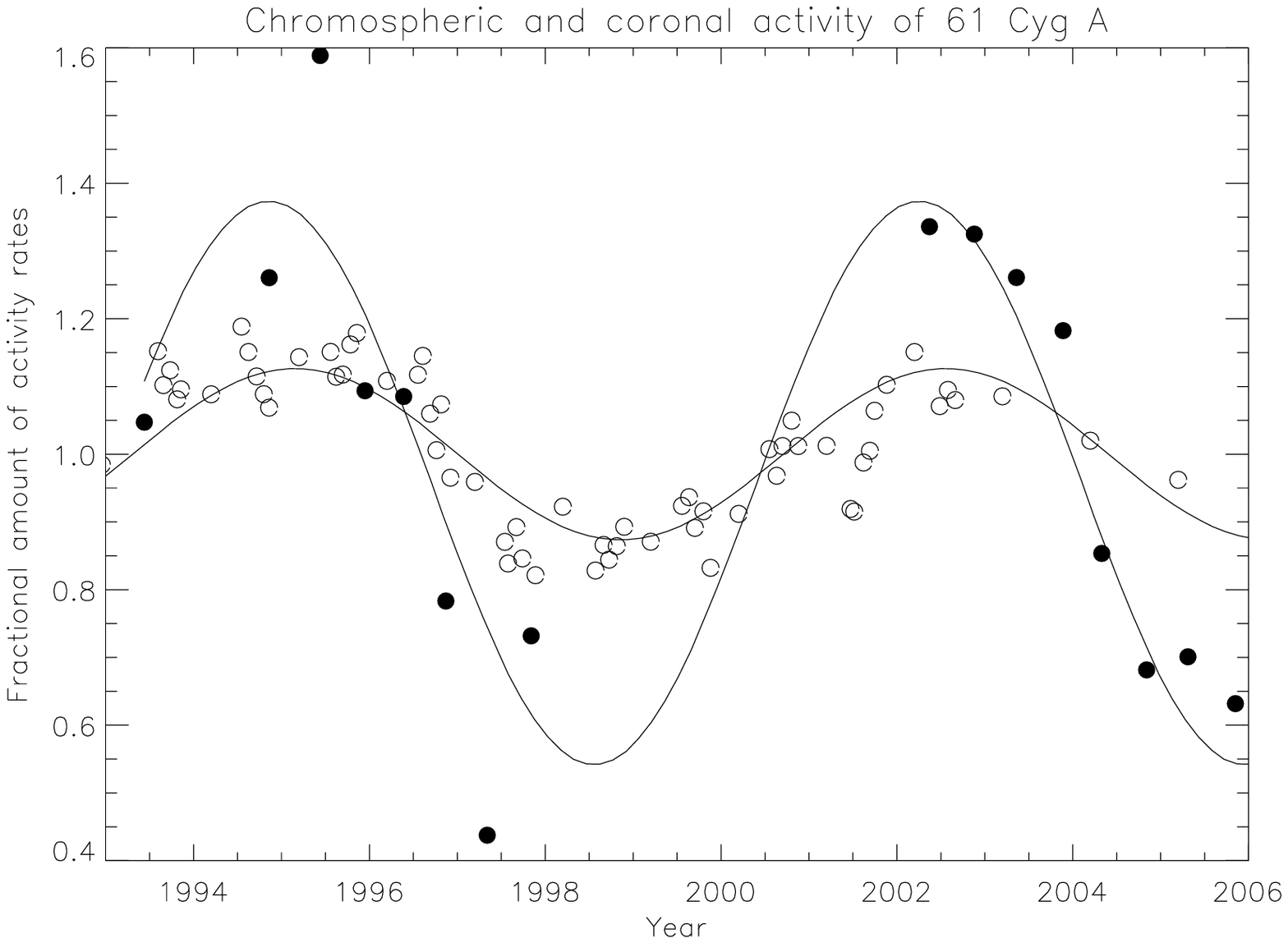}
\includegraphics[width=32mm]{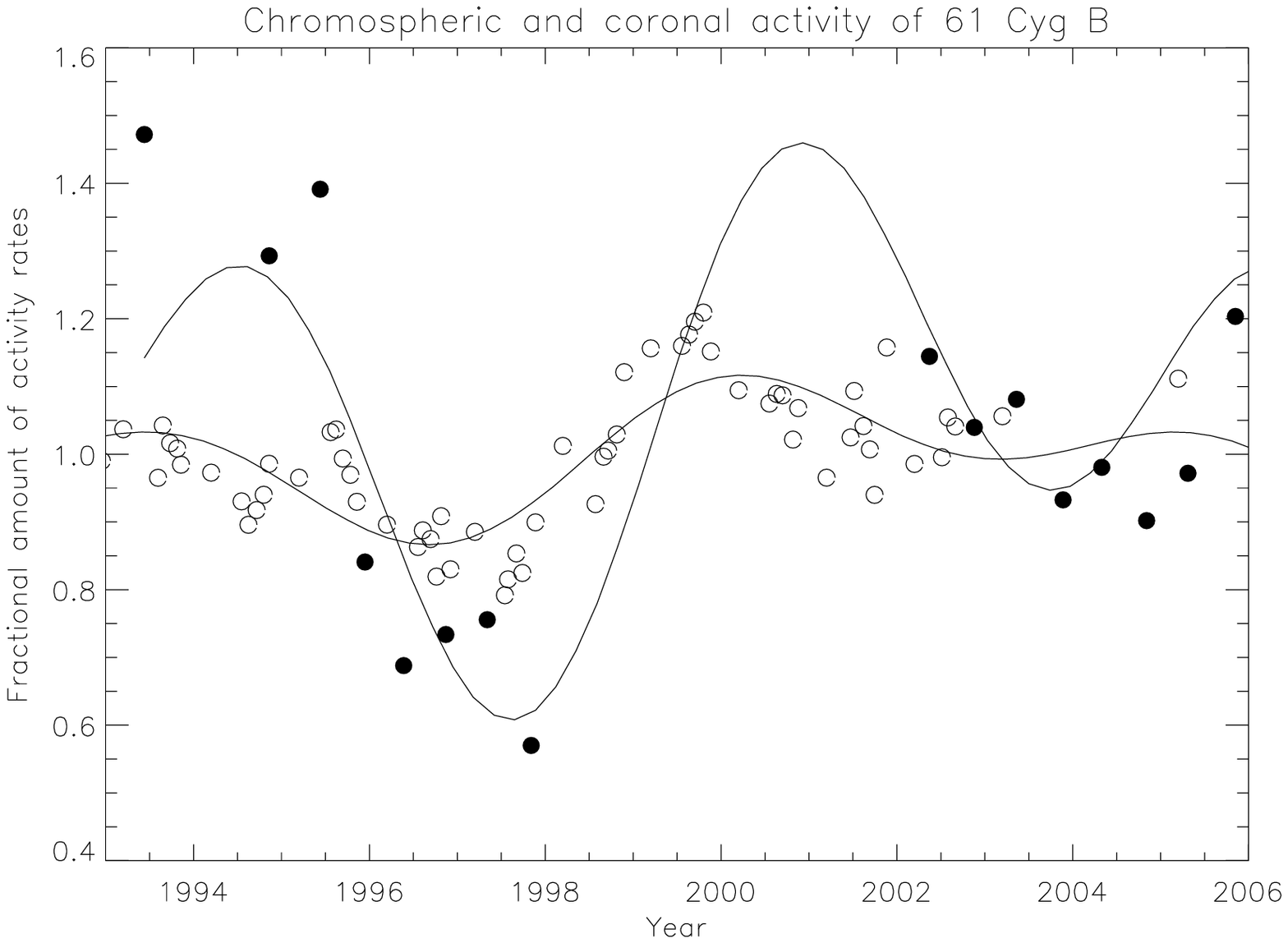}

\vspace*{-0.9cm}
\hspace*{0.8mm}
\includegraphics[width=31.5mm]{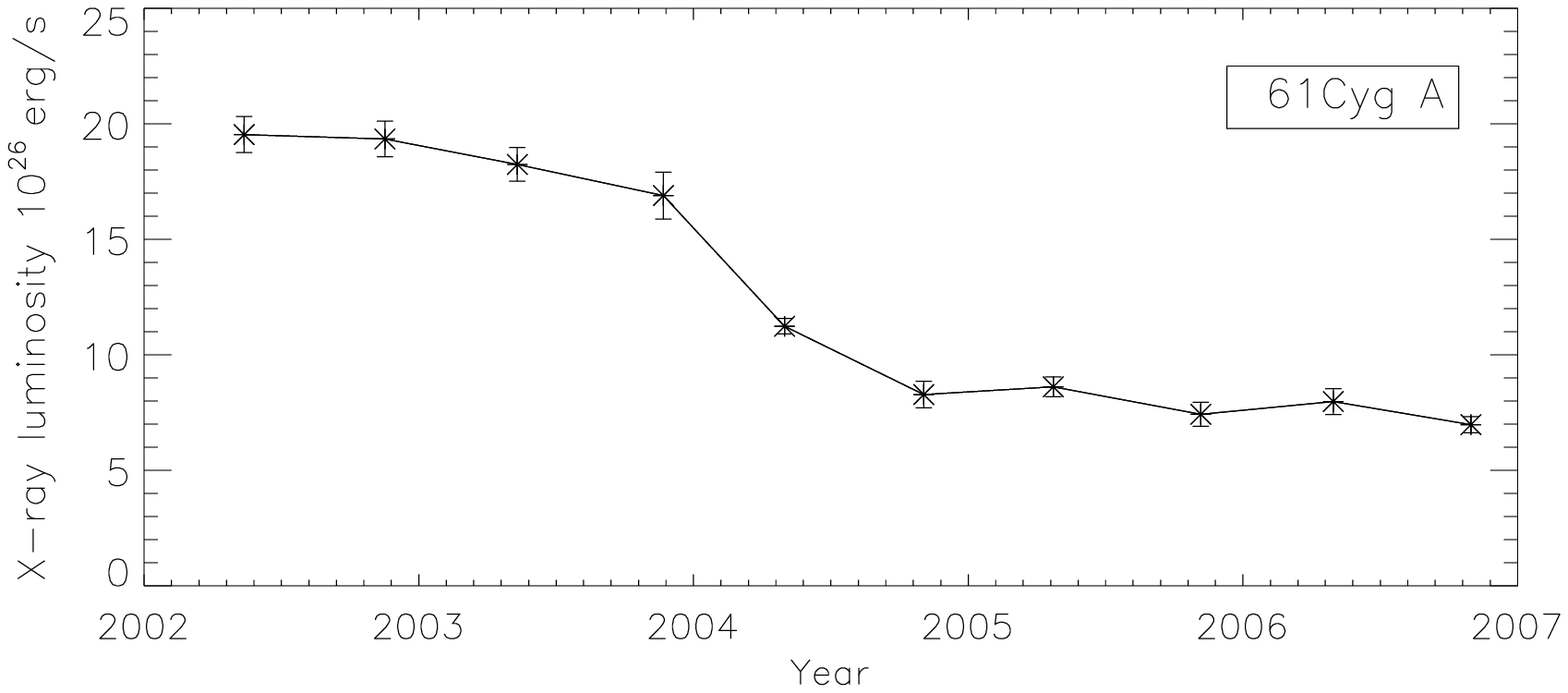}
\includegraphics[width=31.5mm]{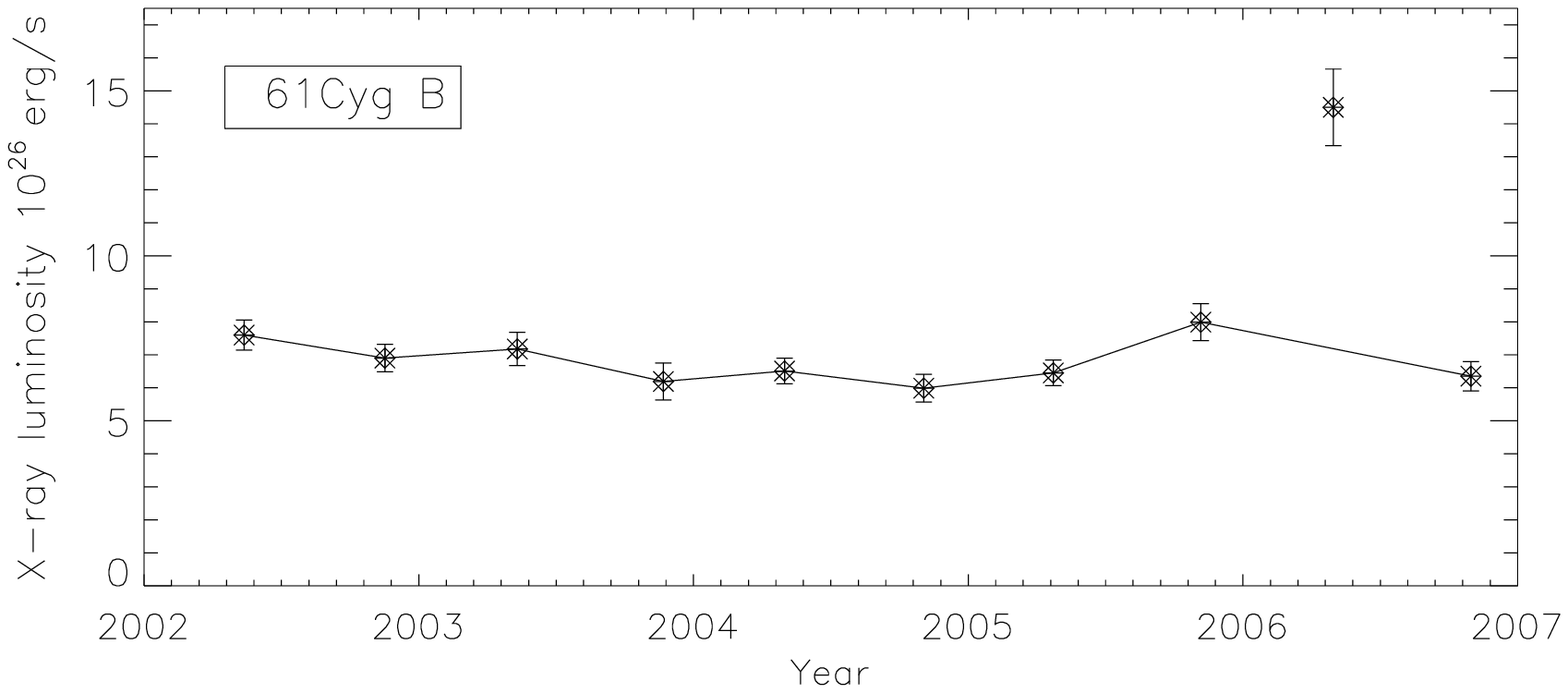}

\caption{\footnotesize Long-term Ca~II H+K measurements~{\it (top)}, comparison between chromospheric (open dots) and coronal activity (full dots) {\it (middle)} 
and X-ray light curves {\it (bottom)} for 61\,Cyg~A and B.}
\label{61cygab}
\end{figure}

A decline in X-ray brightness by a factor of three is observed for 61\,Cyg~A (see Fig.\ref{61cygab}).
Our X-ray measurements are well correlated and in phase with
chromospheric data from the Mt. Wilson project and its follow-up program at Lowell Observatory,
as well as the ROSAT data from the 1990's.
We note, that the cycle amplitude of 61\,Cyg~A is much smaller compared to those of the Sun,
while the $L_{X}/L_{bol}$ ratio and its chromospheric activity index are much higher. 
The chromospheric cycle of 61\,Cyg~B is characterised by more irregular variations as likewise 
seen in its coronal behaviour, where no clear long-term trend is visible in our observations

The spectra of both components are quite similar and their emission measure distributions are
dominated by cool (1\,--\,4\,MK) plasma with variable contributions from a hotter component of about 8\,MK.
The decrease in $L_{\rm X}$ over the course of the activity cycle on 61\,Cyg~A
is mainly caused by a decrease of emission measure accompanied by a moderate decrease in temperature.
The contribution the hot plasma component to the total emission measure
decreases from $\sim$ 15\% at activity maximum to $\sim$ 3\% at minimum.

\section{Alpha Centauri}

\begin{figure}[]
\includegraphics[width=32mm]{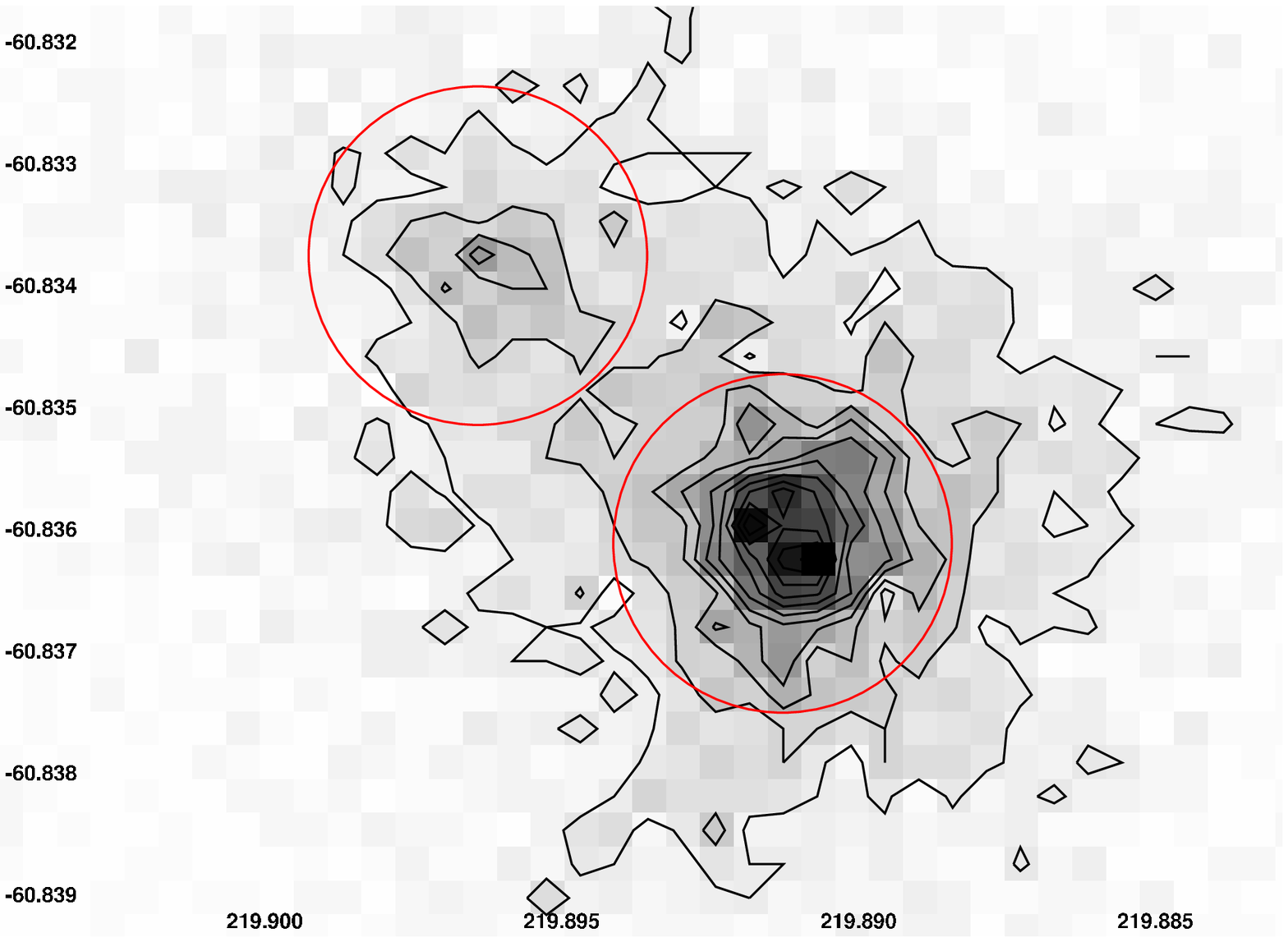}
\includegraphics[width=32mm]{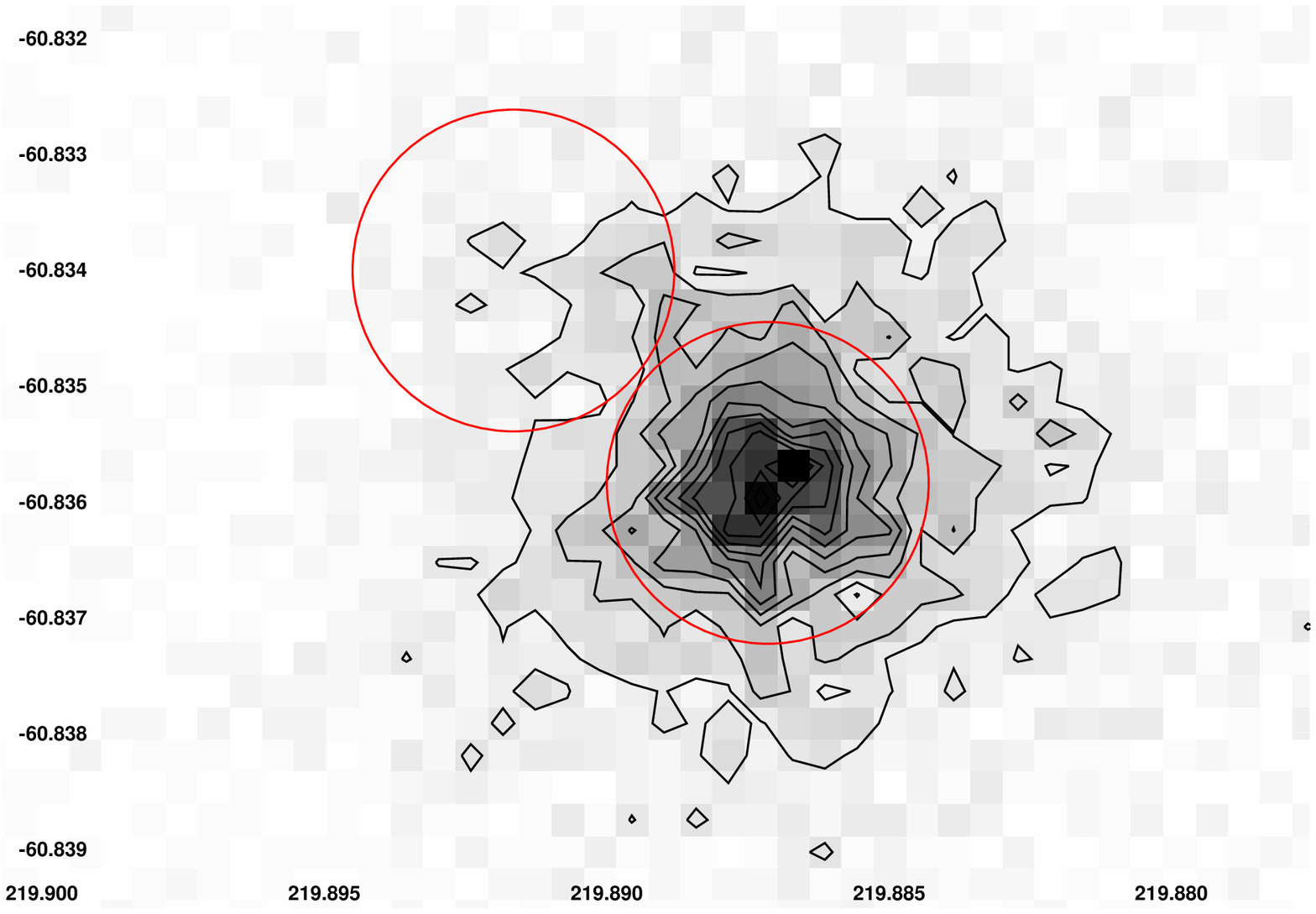}
\caption{\footnotesize 
The Alpha Centauri system observed with MOS1 during March 2003 (left) and February 2005 (right),
overlayed are brightness contours and 5\,\arcsec\, source regions (red) to indicate the proper motion.
The X-ray darkening of $\alpha$\,Cen\,A (upper left) is observed with XMM-Newton for the first time.}
\label{acen}
\end{figure}

\begin{figure}[]
\resizebox{\hsize}{!}{\includegraphics{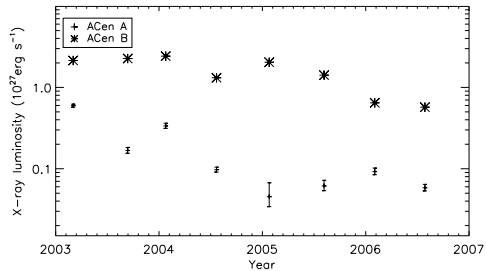}}
\caption{\footnotesize X-ray luminosity (0.2\,--\,2.0\,keV)  for $\alpha$\,Cen~A and B as determined from 
MOS data.}
\label{acenab}
\end{figure}

For $\alpha$\,Cen~A  we find a strong decline in X-ray luminosity by no less than an order of magnitude over 
the time span of a few years (see Figs.\,\ref{acen}, \ref{acenab}), a behaviour that has never been observed before over the last 25 years.
Spectral analysis shows that both stars have a rather cool (1\,--3\,MK) and inactive corona. The darkening of $\alpha$\,Cen~A is again primarily caused by
a strong decrease of emission measure accompanied by a moderate decrease of coronal temperatures.
The absence of long-term chromospheric activity data for these
stars make a definite statement on this point impossible, but recent FUSE observations of transition region lines (e.g. OVI) confirm its darkening.
If real, the activity cycle on $\alpha$\,Cen~A, would then be the first X-ray activity cycle on a true solar analog.
The decline in $L_{\rm X}$ of $\alpha$\,Cen~B may also point to an
X-ray activity cycle, however it has probably to be characterized by
a longer period and/or a smaller modulation.

\section{Conclusions}

The XMM-Newton observations show that coronal activity cycles are present in moderately active solar-like stars.
The observations confirm the existence of a coronal activity cycle for 61\,Cyg~A. This is to our knowledge
the first persistent stellar coronal activity cycle ever detected. 
Its X-ray light curve is well correlated and in phase with ROSAT observation from the 1990s as well as
chromospheric activity measurements. 
We also find a strong decline in X-ray luminosity of the "solar twin" $\alpha$\,Cen~A by no less than an order of magnitude,
probably indicating a coronal activity cycle - however an irregular event cannot be ruled out completely. 
The X-ray amplitude in the more active and hotter corona of 61\,Cyg A is with a factor of three
much smaller than observed for the Sun and $\alpha$\,Cen~A, where it is about one order of magnitude. 
In both cases we find the decrease in $L_{\rm X}$ to be mainly caused by a decrease in emission measure,
accompanied by a moderate cooling of their coronae.
Thus the Sun and our sample stars exhibit very similar behaviour, despite their different
spectral types, activity levels, coronal temperatures and cycle periods. 
This suggests that the underlying mechanisms of activity cycles in low to moderately active cool stars are comparable.

\begin{acknowledgements}
This work is based on observations obtained with XMM-Newton, an ESA science
mission with instruments and contributions directly funded by ESA Member
States and the USA (NASA).
J.R. acknowledges support from DLR under 50OR0105.
\end{acknowledgements}

\bibliographystyle{aa}

\begin{thebibliography}{}
\bibitem[Baliunas et al. (1995)]{bal95}Baliunas S.L., Donahue R.A., Soon W.H., et al. 1995, ApJ, 438, 269
\bibitem[Favata et al. (2004)]{fav04}Favata, F., Micela, G., Baliunas, S.L. et al. 2004, A\&A, 418, L13
\bibitem[Golub et al. (1982)]{gol82}Golub, L., Harnden, F.R.,Jr., Pallacicini, R., et al. 1982, ApJ 253,242
\bibitem[Hempelmann et al. (2003)]{hem03}Hempelmann, A.,Schmitt, J.H.M.M.,Baliunas, S.L., et al. 2003, A\&A, 406, L39 
\bibitem[Hempelmann et al. (2006)]{hem06}Hempelmann,\,A., Robrade,J., Schmitt, J.H.M.M.  et al. 2006, A\&A, 460, 261
\bibitem[Raassen et al. (2003)]{ras03}Raassen, A.J.J., Ness, J.-U., Mewe, R., et al. 2003, A\&A, 400, 671
\bibitem[Robrade et al. (2005)]{rob05}Robrade, J., Schmitt, J.H.M.M. \& Favata, F. 2005, A\&A, 442, 315
\end{thebibliography}

\end{document}